# Accelerating Scientific Publication in Biology

Ronald D. Vale

Dept. of Cellular and Molecular Pharmacology and the Howard Hughes Medical Institute, University of California, San Francisco, CA 94158

Send correspondence to: vale@ucsf.edu

*Abstract*

Scientific publications enable results and ideas to be transmitted throughout the scientific community. The number and type of journal publications also have become the primary criteria used in evaluating career advancement. Our analysis suggests that publication practices have changed considerably in the life sciences over the past thirty years. More experimental data is now required for publication, and the average time required for graduate students to publish their first paper has increased and is approaching the desirable duration of Ph.D. training. Since publication is generally a requirement for career progression, schemes to reduce the time of graduate student and postdoctoral training may be difficult to implement without also considering new mechanisms for accelerating communication of their work. The increasing time to publication also delays potential catalytic effects that ensue when many scientists have access to new information. The time has come for life scientists, funding agencies, and publishers to discuss how to communicate new findings in a way that best serves the interests of the public and the scientific community.

Most biologists have become frustrated with the current state of scientific publishing. Attention has been drawn to flaws in using journal impact factors for evaluating scientific merit (1), the hypercompetitive environment created by scientists seeking to publish their work in the top journals (2), and the extensive revisions required by reviewers and editors (3, 4). In this Perspective, I wish to focus on another issue that has received less attention– the increasing amount of data and time required to publish a paper.

As a consumer of scientific literature, I enjoy reading the comprehensive scientific studies that are being published today. However, the foundation of today's data-rich articles is acquired at a cost, which is the time that graduate students and postdoctoral fellows spend in collecting and analyzing data. Indeed, as I will discuss later, the length of time required to produce and then publish a scientific work is likely impacting the duration and quality of Ph.D. and postdoctoral training. Furthermore, as laboratories wait to accumulate more experimental data before they feel that a benchmark for publication is met, crucial results are being sequestered from the scientific community for longer periods of time. In this Perspective, I will argue that creating new outlets for faster and more nimble scientific communication could have positive outcomes on professional training, catalyzing scientific progress, and improving the culture of communication within the life sciences as a whole.

***A trend toward increasing data required for publication***

Many senior scientists feel that the amount of data required for publication has increased over their careers (for example, see ref. 4). But is this actually true? Quantifying the amount of experimental data in a publication is non-trivial, as data can take many different forms and varies in the amount of time required for its acquisition. Furthermore, comparing the amount of data in contemporary versus prior papers is difficult. For example, the time required to obtain certain types of information has decreased; as an extreme example, sequencing an entire genome now requires less time than cloning and sequencing a single gene 40 years ago. However, scientists always push technical limits, and many of the experiments performed today also are difficult and require a long time to master and execute. Thus, I would argue that truly informative experimental data are not vastly easier to obtain now than in the past. Practices in data inclusion, however, may have changed; for example, experiments previously described as "data not shown" would now likely be included in a supplemental figure, and figures also are easier to prepare now with computer programs compared with more cumbersome manual methods in the past.

With the above caveats noted, I sought to compare the amount of experimental information presented in biology papers published Cell, Nature (biology only), and the Journal of Cell Biology (JCB, operated by editors from the scientific community) from the first six months of 1984 and of 2014. The number of papers published by Cell remained approximately the same, decreased slightly for Nature, and dropped in half for the JCB in 2014 compared with 1984 (Fig. 1A). The average number of figures in the print version of papers did not change significantly, as journal guidelines have remained largely the same between these two time periods (Fig. 1B). However, during this thirty-year span, the number of experimental panels contained within the print version of the paper rose dramatically by 2-4 fold (Fig. 1C; see Fig. S1 for the breakdown of short and long format papers in Nature and JCB). Separate labeled panels do not always constitute distinct experiments, and figure labeling styles might have changed in past thirty years. To examine this point, panels in Cell and Nature were scored as to whether they contain distinct pieces of data or were derived from the same experiment (see Supporting Information Methods and Fig. S2). The number of distinct data sets was approximately two-thirds of the number of labeled panels, and this ratio did not change substantially between 1984 and 2014 for either Cell or Nature. Thus, the fold-increase in panel number appears to reflect a true increase in the amount of data in the print version between 1984 and 2014. The increase in the amount of data per paper is even more substantial when supplemental information, which began to appear ~1997, is taken into consideration (Fig. 1B, C). In particular, the number of supplemental figures and their panels were comparable to (Cell) or exceeded (Nature) those that were published in the print version (Fig. 1C). Consistent with this trend of more data and the likely use of more diverse and complex techniques, today's papers in Cell, Nature and JCB have 2-4 fold more authors than those from 1984 (Fig. 1D). However, enlisting more authors is probably not the sole mechanism for acquiring the additional data needed for contemporary papers.  As will be discussed later, it also appears to take a longer period of time to publish a paper now than in the past.

*Factors driving an increasing amount of data per publication*

What factors have driven the increasing amount of data per publication over the past few decades? One likely factor is supply and demand- more scientists are competing for the same or less real estate (space in top journals, Fig. 1A) compared to thirty years ago. Over the past 30 years, the US scientific workforce (e.g. postdoctoral fellows and graduate students) has increased by almost three-fold (5, 6), fueled, in part, by the doubling of the NIH budget between 1998-2003. In addition to the US, many other countries recently have expanded their life science research programs. From 1999 to 2005, publications from US labs increased only 3.6% annually, while those from China increased 38.9% (7). Thus, with more scientists desiring high-profile publications for their grants and promotions, the elite journals can set a higher bar for what they accept.  A "high impact" result constitutes one important criterion for publication. However, a second and increasingly important benchmark is having a very well developed or "mature" research story, which effectively translates into more experiments and more data.  A whole genome screen followed by a mouse model to understand the physiological functions of one of the gene hits as well as additional structural work to understand the mechanism might be what is needed to seal the deal for acceptance. Reviewers, in turn, fall in line with the escalating expectations and continually reset their own benchmarks of "what it takes" to get into a particular journal. With these market forces at work and a positive feedback loop between journal editors and reviewers, the expectations for publication have ratcheted up insidiously over the past few decades.

In addition to the time required to obtain the data for submission, the review process itself typically adds new demands for more data before the work can be officially accepted for publication. If one is fortunate enough to have the paper sent out for review, then three referee reports are commonplace these days. Frequently, each referee requests additional experiments.  Many of our own papers have been significantly improved by experiments suggested through peer review. However, many suggested experiments are unnecessary, and sometimes the requested work is so extensive that it constitutes a separate study onto itself. Furthermore, it is not easy to "say no" to referee-suggested experiments or a journal request to curtail the discussion. After all, the journal editor will have another revised paper on his/her desk where all of the referees are completely satisfied. Thus, authors feel as though they are held hostage, fearful that their paper will not be accepted if they do not comply with most, if not all, of the requests.

While the elite journals are important driving forces in the scientific market place, the trend towards more data is felt throughout the publication ecosystem. One reason is that non-elite journals want to improve their status, and, as a consequence, strive to be selective and seek more mature stories. This is perhaps why JCB accepts fewer papers now than it did in the 1980s (Fig. 1A). Second, scientists feel pressured to aim high and acquire the data that they *think* will be needed for publication in an elite journal. But alas, when it comes time for journal courtship, they find their work editorially rejected not once, but thrice, and then eventually publish their large body of work in a lower tier journal. It is not easy to obtain information on journal rejections from the 1980s, although I speculate that the frequency has increased considerably in the past thirty years. Thus, in addition to the time invested in acquiring data, the time spent in finding a home for a paper through sequential journal submissions also significantly delays the transmission of results to the scientific community.

### *What is a minimal unit for publication?*

Most scientific papers, now and in the past, usually have one or two key findings. But with the trend towards publishing more mature scientific stories, it has become harder to publish just a key initial finding or a bold hypothesis.

Let's consider the Watson and Crick publications, perhaps the most famous in modern biology, and imagine how they might fare in today's publishing environment. Many people may be unaware that Watson and Crick published not one but two papers on DNA in Nature in successive months. The first paper published on April 25, 1953 described a structural model for the DNA double helix (8). Despite having a single figure (a model figure without data), it was listed as an "Article" rather than a "Letter", based upon the magnitude of the idea. The first Watson/Crick paper was accompanied by two other Articles on the X-ray diffraction pattern of DNA; the paper by Maurice Wilkins had two figures (9) and the one by Rosalind Franklin displayed a single figure (10). The second Watson/Crick Nature paper (also an Article published on May 30) was entitled "Genetic Implications of the Structure of Deoxyribonucleic Acid". It described, without any data, a hypothesis for the hydrogen bonding of the "Watson-Crick" base pairs and speculated how the two DNA strands might each provide a template for the replication of genetic information (11). Several months later, Wilkins and Franklin each independently published second Nature Articles describing more complete analyses of the structure of DNA (12, 13). Thus, the story of DNA, like a Charles Dickens novel, came out in installments. Furthermore, it also should be emphasized that the Watson and Crick model was speculative, particularly with regard to the process of DNA replication. As a result, the revolutionary ideas of Watson and Crick were not instantly accepted and their implications were not widely understood by the scientific community at the time of publication. Experimental evidence for the unwinding of the DNA strands and semi-conservative replication was published in 1958 by Meselson and Stahl (14), and this placed the Watson and Crick model for replication on a solid footing.

Somewhat tongue-in-cheek, let's imagine a contemporary editorial decision on the 1953 Watson and Crick papers (in reality, these papers were not peer reviewed; see Nature's recollection of the publication process (15)):

*"Dear Jim and Francis:*

*Your two papers have now been seen by three referees. Based upon these reviews, I regret to say that we cannot offer publication at this time. While your model is very appealing, referee 3 finds that it is somewhat speculative and premature for publication. Indeed, your model proposing a semi-conservative replication of DNA raises many obvious questions. As two of the referees point out, it should be possible to determine experimentally if the two strands can separate and serve as templates. This would address referee 3's concern that strand separation is not feasible thermodynamically. I regret to say that without such experimental evidence, we will not be able to publish your work in Nature and suggest publication in a more specialized journal. Should you be able to furnish more direct experimental evidence, we would be willing to reconsider such a revised paper. Naturally we would need to consult our referees once again. Furthermore, since space in our journal is at a*

*premium, if you do decide to resubmit, then we recommend that you combine your two submitted papers into a single and more cohesive Article, potentially including the X-ray studies of your colleagues at Cambridge. Thank you again for submitting your papers to Nature. I am sure that this revision will delay your Nobel Prize and the discovery of the genetic code by only one or two years."*

A discovery emerging in closely spaced installments was not unique to DNA. The molecular mechanism underlying familial hypercholesterolemia was unraveled in three key papers by Brown and Goldstein between 1973-1974, each of which solved a piece of the puzzle (16-18). Similarly, the discoveries of ubiquitination and protein degradation by Hershko, Ciechanover, and Rose emerged in three papers in 1979-1980 (19-21). Studies on the mechanism of axonal transport by myself, Schnapp, Reese and Sheetz (covering work from 1983-1985) were published in five papers in 1985 (22-26). In all of the above examples, the information could have been delayed and compacted into fewer publications, as no doubt would occur today. However, by unfolding these breakthroughs in a series of papers, the progression of results could be quickly disseminated to the scientific community, the value of which will be discussed in the next section.

Today, two opposing factors come into play in deciding when to publish a paper. On one hand, scientists want to get their work published as fast as possible, both for advancing their careers as well as claiming priority for their discovery and avoiding getting "scooped". However, publishing in a top journal has become an equally compelling consideration for many scientists, and this latter factor can tip the balance towards delaying submission until more experimental data can be obtained.

### *Consequences on the exchange of information within the scientific community*

The "comprehensive" paper enables authors to build a convincing argument for their hypothesis. Indeed, the Watson/Crick model combined with the Meselson/Stahl experiment would have constituted an amazing paper that would have immediately convinced everyone in the field. However, there is also merit in getting new ideas and key experiments published with reasonable speed, even if they are incomplete. Once in the public domain, the collective power of the scientific enterprise can take effect and the ideas can be tested and advanced further, not only by the original researchers but by other investigators as well. Once results are published, other scientists can see connections with their own work, perform new experiments that the original investigators might never do, and also emerge with new ideas. Overall, putting new results and ideas in the public domain is good for science and serves the mission of the funding agencies that seek to advance research overall.

The protracted and uncertain nature of the publication process also may be affecting the exchange of information at scientific meetings. Students and postdocs, although eager to have the chance to present their work, have become increasingly wary about sharing their unpublished data at scientific meetings. As a result, scientific meetings are becoming increasingly filled with recently published or soon-to-be published results, rather than exciting work in progress.

### *Consequences for Training*

In 1990, the average age at which scientists received their first R01 NIH grant was less than 38 years; in 2013, that same milestone was reached at an average age of over 45 years (27). This trend is of great concern for many obvious reasons (2, 27), including the fact that it is making a career in biomedical research less attractive to young people (28). In an attempt to reverse this trend, efforts are now being made to accelerate the career track of young scientists. Many graduate schools require regular thesis committee meetings to promote timely graduation, and a recent Perspective in PNAS suggests limiting funds for graduate training to 5 years (29). Some institutions and granting agencies limit the length of postdoctoral training to 5 years, which is also strongly recommended by a recent National Research Council report (30) and others (29). In addition, new grant schemes, such as the NIH K99, seek to promote the transition of postdoctoral fellows to junior faculty positions. All of these measures are worthy, but for them to succeed in reducing training time, they must be accompanied by changes in the publication system. Placing term limits on graduate and postdoc training would be a perfect solution if PIs were always responsible for keeping their trainees for too long in their laboratories. While this no doubt occurs, graduate students and postdocs also are asking their PIs if they can stay for a longer period of time. To understand why this is happening, one has to appreciate the connection between publication and career advancement.

Scientific papers are required for obtaining a job, a promotion, or a grant, and thus have become a primary currency for professional advancement. Furthermore, papers in elite journals have become particularly valuable in the career marketplace. Graduate students and postdocs understand the "paper economy", and they want to publish as many papers as possible and ideally publish a paper in Cell, Science or Nature.

But it seems as though publishing many papers and being published in elite journals is harder now than it was in the past. I examined the publication records for Ph.D. students at University of California San Francisco (UCSF) who graduated in the 1980s (n = 71) versus those that graduated in the past three years (n= 104; Table 1; Fig. S3 and S4). The average time for acquiring a Ph.D. increased slightly between the past (5.7 years) and current (6.3 years) student groups; these times to degree are largely consistent with national trends (5, 29). However, even though the contemporary group of graduate students was in school for one-half year longer, they published fewer first/second author papers and published much less frequently in the three most prestigious journals. Consistent with the notion of more data being required for publication, the contemporary students also took an additional 1.3 years, on average, to publish their first, first-author paper compared with students from the 1980s. Strikingly, the average time to a first author publication for the current cohort (6 years for students who publish) is just below the average time of their graduation (6.3 year) and at the desired upper boundary for training in these graduate programs (6 years or less). These general trends also are apparent when comparing the top 1/3rd of students with the best publication records, suggesting that the differences cannot be explained by admitting a pool of less capable students now than in the past (Table 1). UCSF also remains a highly sought-after graduate school, and its reputation has gotten stronger since the 1980s. This type of analysis should be extended to larger numbers of students from many different universities, but these preliminary data suggest that it has become harder for graduate students to publish.

The increasing time to publication poses difficulties in reaching milestones for career advancement. Graduate students often need to apply for a postdoctoral position 9-12 months prior to graduation and thesis committees frequently recommend having a first-author paper accepted for publication prior to initiating the application process. Postdocs seeking a job or grant support face a similar predicament. For example, let's consider the timing of the highly sought-after NIH K99 Pathway to Independence Award, which provides 1-2 years of postdoctoral training and 3 years of independent support. The postdoc likely requires 2 months to write a successful grant and then it can take 9 months from submission to the time when funding is received. Importantly, a K99 grant will be considered much more competitive if the postdoc has a prior publication; a "manuscript in submission" cannot be listed in an NIH grant application. If it takes a postdoc three years to have a paper accepted before submitting a competitive K99 application (often a best case scenario), then a talented young scientist will spend ~5-6 years in a postdoc before getting a job (three years to publish a paper, an additional year from grant writing to funding, followed by a ~1-2 year training period). In summary, the ability of thesis, grant, and job committees to access a formal and publicly accessible paper could accelerate career transitions towards the end of graduate and postdoctoral training.

Providing young scientists with more opportunities to publish also has other advantages for training. Preparing and publishing a scientific paper is a critical part of the apprenticeship of becoming a scientist. This experience not only promotes skills in writing, but also in organizing experimental data and learning how to convey ideas effectively. The process of completing a scientific paper also teaches young scientists how to be more efficient in planning and executing experiments in their future projects. However, with the increasing time involved in acquiring data and publishing, young scientists get fewer chances to write papers and thus arguably are less well trained in these skills than trainees in the past (Table 1). Furthermore, if a critical study reaches the point of publication after 4-5 years of work, all too often the PI, who has more experience, takes over the process of writing from a graduate student or postdoc. In such cases, neither the young scientist nor the PI are willing to take chances with the paper being accepted in today's competitive publication environment.

Another value of publishing earlier is that it allows a graduate student or a postdoc to explore more options for utilizing their remaining training period. Rather than myopically focusing on getting their one paper out, trainees can decide whether they want expand their first study, move on to another research question, or spend some time pursuing additional career training (e.g. teaching).

***Possible solutions for accelerating communication***

New journals and publishing platforms have recently introduced several interesting innovations, including providing immediate open access to publications (which PLoS One is doing on a large scale) and reforming the process and transparency of peer review (e.g. eLife and F1000 Research). The above efforts should be applauded. However, creating more new journals, which are expensive to operate and must struggle to compete for good manuscripts, is unlikely to constitute the transformative solution needed for accelerating scientific communication. A mechanism that has the potential for transformative change must: 1) operate on a large scale (i.e. hundreds of thousands of papers per year rather than hundreds), 2) succeed in capturing the very best work in the field, 3) be able to launch and co-exist with existing journals, and 4) be cost-effective and be possible to implement on a time scale of years rather than decades.

*Lessons from the Physics Community: Should Biologists Adopt an Internet Pre-Print System?*

A mechanism for accelerating scientific communication that meets the above criteria has been developed already by the physical science community. Physicists, mathematicians, and computer scientists typically deposit their scientific manuscripts prior to journal publication in an open access e-print service called arXiv (pronounced "archive"), which was founded by Paul Ginsparg and is now operated by Cornell Library. At first created for the high energy physics community, arXiv usage has spread over time to other sectors of physics, mathematics, computer science, and quantitative biology. This repository of electronic pre-prints is searchable, and many physicists have developed a habit of checking for alerts from arXiv first thing in the morning. Generally, although not always, a paper uploaded onto arXiv is then submitted to a journal. Importantly, the public disclosure through arXiv is accepted by the physical science/mathematics community as priority for a discovery, and an arXiv posting is acceptable as a reference in a journal, book or grant application. After the original paper is posted in arXiv, new versions can be uploaded, for example after a paper has been revised through the journal review process or in response to other comments received by the community. However, earlier versions of the paper are retained and the nature of the changes are indicated in revised uploads.

ArXiv evolved from a common practice in the physics community, beginning several decades ago, of mailing unpublished manuscripts to colleagues in the field. This also was more common in the early years of molecular biology, a famous example being Watson and Crick obtaining a pre-print from Linus Pauling that proposed the erroneous triple helix model of DNA. As technology evolved, mail turned to email, and physicists sent their manuscripts to colleagues by this electronic route. With the development of the internet, physicists rallied around the formation of a pre-print server, and arXiv was established in 1991. From its inception through January 2015, one million papers have been submitted to arXiv. In 2013 alone, arXiv papers were downloaded 67 million times. Differing from the bulk of work in biology, arXiv contains many purely theoretical papers. However landmark experimental studies also are routinely disseminated first on arXiv, a recent example being the discovery of the Higgs boson.

Would a centralized, open access, and widely used pre-print repository be sensible for biologists, as it has been for physicists? Harold Varmus advocated for such a system (termed E-biomed) in 1999 when he was director of the NIH (31) and others have more recently echoed benefits (32). Currently, there are a few pre-print servers specifically for biology, including bioRxiv.org (launched in 2013 by the non-profit Cold Spring Harbor Press) as well as PeerJ and F1000 Research, for-profit companies that also offer platforms for peer review. However, pre-prints in biology have not achieved a critical mass for take-off. Last year, for example, bioRxiv received 888 pre-prints compared to 97,517 for arXiv, even though many more papers are published in the life sciences.

Having never used a pre-print server myself, I tried the experiment of submitting this Perspective to bioRxiv and PNAS on the same day (July 10, 2015); after initial screening, the article was posted as a PDF on bioRxiv on July 11 (33). Fig. 2 shows the number of views of the bioRxiv article and social media exchanges ("tweets") from the time of pre-print posting until the receipt of two peer reviews and an editorial decision from PNAS (August 21). The data show the pre-print reached a large audience (views of the abstract were over twice that of whole article) and also reveal how social media can drive viewership. Importantly, even prior to the receipt of two anonymous referee reports, I received extensive feedback on the article through comments posted on bioRxiv, direct emails from readers, and numerous personal discussions. Such feedback helped me to formulate a set of the pros, cons, and uncertainties surrounding pre-prints, as discussed below (for a more extensive discussion of these issues, see the Q&A in the Supporting Information).

The Pros: Fast, Free and Feasible

1) Submission to a pre-print repository allows a paper to be seen and evaluated by colleagues and search/grant committees immediately after its completion. This could enable trainees to apply for postdoctoral positions, grants, or jobs earlier than waiting for the final journal publication. It also allows independent investigators to transmit their latest work in a reliable manner to grant review committees, without an unknown delay imposed by the journal publication process. A recent study of several journals found an average delay of ~7 months from acceptance to publication (34), but some journals take longer (34) and this time does not take into account journal rejections and the increasingly prevalent need to "shop" for a journal that will publish the work.

2) A primary objective of a pre-print repository is to transmit scientific results more rapidly to the scientific community, which should appeal to funding agencies whose main objective is to catalyze new discoveries overall. Furthermore, authors can receive faster and broader feedback on their work than occurs through peer review, as I have discussed as a case in point with this article (also see an experience from a junior faculty member in the Q&A, SI).

3) If widely adopted, a pre-print repository (which acts as an umbrella to collect all scientific work and is not associated with any specific journal) could have the welcoming effect of having colleagues read and evaluate scientific work *before it has been branded with a journal name*. For grants, jobs and awards, physicists will read and evaluate science posted on arXiv. The life science community needs to return to a culture of evaluating scientific merit from reading manuscripts, rather than basing judgment on where papers are published.

4) A pre-print repository is good value in terms of impact and information transferred per dollar spent. Compared to operating a journal, the cost of running arXiv is low, with most of its operating costs covered from modest subscription payments from 175 institutions and a matching grant from the Simons Foundation. Unlike a journal, submissions to arXiv (and currently bioRxiv) are free.

5) Future innovations and experiments in peer-to-peer commentary and evaluation could be built around an open pre-print server. Indeed, such communications might provide additional information and thus aid journal-based peer review (described below).

6) A pre-print server for biology represents a *feasible* action item, since the physicists/mathematicians have proof-of-principle that this system works and arXiv has co-existed with journals, with each providing different services in science communication (see Q&A in SI).

Cons: Lack of peer review and information overload

1) The lack of peer review might invite lower quality or irreproducible data to be disseminated. While a risk particularly for certain medical research (see Q&A in SI), several factors mitigate such concerns. First, arXiv and bioRxiv each have an initial screening mechanism that helps to eliminate overtly "unscientific" articles. Second, the major factor for ensuring quality is that the reputation of the investigator is at stake, and achieving a good reputation within the community is a primary motivating factor for scientists. Indeed, a pre-print submission is immediately visible to the entire community, whereas a journal submission is seen confidentially by only a couple of referees. Thus, posting of a poor quality paper on a pre-print server will be widely visible and reflect poorly on the investigator and his/her lab. Scientists take pride in their work and will be guided by their own internal standards in deciding when their work is ready to be released to the community. Third, the paper can receive input (as this article has) from more than 2-3 referees, which could help authors correct flawed experiments/statements and help produce a better final product published in the journal. Fourth, peer review by journals, while helpful, is certainly not a fool-proof mechanism for identifying problems or eliminating scientific irreproducibility, especially since the referees' first task is to assess whether the work is "exciting enough" rather than "accurate enough". If a recent fictitious method for preparing pluripotent stem cells (35) had first surfaced as a pre-print, many scientists would have likely noted its flaws well before journal publication. Thus, the buyer always must beware and exercise appropriate judgment for scientific quality, regardless of whether a study appears in an elite journal or an electronic pre-print server. In addition, one could imagine an option of incorporating author-initiated peer evaluations as part of a pre-print, which most scientists do informally before submitting their work to a journal and is not unlike the mechanism by which National Academy of Science members submit papers to the PNAS.

2) Pre-prints could expand the problem of information overload in biology by opening the door to less interesting reports that are not being published by journals. While this could be true, certain "unpublishable" studies, such as a negative result or whether a prior finding can be reproduced, might provide useful information to some scientists. Furthermore, scientists are already living in a world of information overload. Rather than suppressing pre-prints, the answer may lie in better search filters such as key words, colleagues of interest, social media cues, and potentially even other measures of validation (such as whether the work was supported by a grant from NIH, NSF, or other major agencies).

Uncertainties: culture, priority, and government and journal support

If the pros seem attractive and the cons manageable, then why are pre-prints not being used by biologists? One reason is that most biologists simply don't know about pre-print servers. But there are other reasons as well. Many believe that biology has a different culture from physics, which will make it impossible for the success of arXiv to be extended into biology. "Culture" refers to the moral fabric of the community- how credit for a discovery is assigned, how information is shared, and how a scientist's work is evaluated. Currently, many issues regarding pre-prints, which are clear for physicists, are clouded by uncertainty in the biology community (see also Q&A, SI). In the fast moving world of experimental biology, will a pre-print publication result in an increased risk of losing credit and getting "scooped"? Will a pre-print put a journal submission at risk for automatic rejection. Will a pre-print be recognized by grant agencies, thesis committees, etc.? These uncertainties create considerable barriers to use of pre-prints in the biology community. The following leadership and policy changes could eliminate these barriers:

1) Pre-prints become accepted as evidence for establishing priority of a discovery, as is true in physics.

2) Pre-prints become accepted as evidence of productivity in grant applications. Currently, NIH only allows listing of accepted peer-reviewed papers in a grant. However, grant reviewers are "peer reviewers" and should be able to judge the quality of a scientist's most recent work in the form of a pre-print.

3) Pre-prints become accepted by life science journals. Currently, many journals (Science, Nature, eLIFE, PNAS, others) allow prior pre-print submissions; however, some journals still have ambiguous policies, which constitutes an overall deterrent.

*Help from the journals: creating a new "Key Finding" format*

A pre-print server provides a solution for improving the ease and speed of communicating a paper, but it does not necessarily address the escalating amount of data needed for publications in journals (Fig. 1). Here, journals themselves could take the lead. Many journals now have "short" communications (e.g. Nature *Letters*, Science *Reports*, J. Cell Biology *Reports*, Current Biology *Reports*). However, their guidelines have primarily curtailed the number of words rather than the amount of data, as researchers have found creative ways of stuffing more and more into the allowable number of figures and supplemental online material (noting the obvious element of irony, please see supplemental Fig. S1 for the amount of data included in Nature *Letters* and JCB *Reports*). It is worthwhile considering introducing a new journal format whose focus is on limiting data more than text. One could imagine a format limited to 8 panels arranged in up to 4 figures and with no Supplemental Data. One of the figures could be identified as the "Key Finding", with a text box describing why it contains the cornerstone result of the article. Is it possible to convey good science in such a restricted format? It was possible 30 or more years ago (this idea is effectively the Nature Letter or Science Report of the past), so it should be now. Creating a new format has the potential of permeating throughout the publishing world, like cover art, commentaries, etc., provided that it is popular among authors and readers.

**Conclusions**

We may be approaching a breaking point in the publication process in the life sciences. The analysis of graduate students presented here suggests that the average time to first author publication has ratcheted upwards and is now approaching the length of Ph.D. training. Furthermore, the strong desire of investigators and their trainees to publish in high profile journals, the requirements of US graduate programs (implicit or explicit) for Ph.D. candidates to publish a first-author paper, the inability to include not-yet-accepted manuscripts in grant applications, and the hopes of federal agencies to shorten PhD/postdoc training are all coming into conflict with the ground realities of the present day scientific communication system. In addition to scientific training, important elements of scientific culture also stand to gain from improving the practices and timing of publication, including better evaluation practices for promotion and regaining an open atmosphere of communicating unpublished results at scientific meetings.

Changing the status quo appears daunting if not impossible, particularly to many young scientists who feel frustrated by the present publication system. It is easy to assign the fault to the journals, but such blame is misplaced and diverts attention from where the lion's share of the responsibility lies—in our own life sciences community. As scientists, we need to define our culture and take ownership in developing a system for communicating research results that best suits our needs as well as the needs of the public. We have not done so, at least not yet. Optimistically, change can happen if our community sets its mind to the task, recognizing that universal consensus may not be achievable and that certain subfields of biology will likely embrace new ideas more readily than others. Young scientists, who have grown up in a culture of sharing information on the internet, also may embrace a new opportunity, if it is presented to them.

As is often the case, it is easier to articulate the problem than derive an effective solution. One idea discussed here for accelerating publication in the life sciences is the wide-spread adoption of electronic pre-prints. Mechanisms for submitting pre-prints already exist; however, with everyone standing at the shore and very few people willing to jump in, the water looks cold and uninviting. Thus, a challenge for this idea becomes changing behavior on a massive scale, which first requires removing barriers and providing better incentives for pre-print publishing; only then can the experiment be done properly of establishing whether pre-prints serve the needs of biologists. Others may feel that reform of the existing journal system (better and more transparent reviewing, better evaluation metrics) might suffice without resorting to a pre-print server or other new model. But how effective will these reforms be without implementing new incentives for currently overwhelmed scientific referees and will they be sufficient to truly change the "daily lives" of graduate students and postdoctoral fellows? Others feel that journals and pre-prints are both arcane and developing an entirely new system is needed. To discuss and debate these issues, it may be an opportune time to hold a meeting of major stakeholders (junior and senior scientists, funding agencies, scientific societies, philanthropists, and journal editors) *specifically* to discuss the issue of how to accelerate the communication of scientific results in biology. The most important stakeholder in this discussion is the National Institute of Health, which has already greatly influenced publication practices by requiring its grantees to abide by public access policies. Since the NIH is deeply interested in 1) promoting public good by catalyzing research discoveries, a process that is facilitated by rapid access to scientific results, and 2) advancing the career paths of its trainees, the topic of accelerating scientific communication should be of great interest to them. Indeed, everyone will likely step into the water together with new pre-publication and/or publication practices if the NIH determines that it serves the greater good of the scientific community and the nation's research agenda. Through thoughtful discussion, engagement and action, our system of scientific communication can be guided to meet the current needs, challenges and exciting opportunities in the life sciences.

## Acknowledgments

I would like to thank Walter Huynh, Courtney Schroeder, and Phoebe Grigg for their considerable help with the analyses of publications presented in this paper.  I also thank Ron Germain, Satyajit Mayor, Richard Sever, and Harold Varmus for their detailed comments on the initial manuscript, and the many other individuals who commented on the article after it appeared on bioRxiv.

16. Brown, M.S., Dana, S.E., and Goldstein, J.L. (1973). Regulation of 3-Hydroxy-3-methylglutaryl coenzyme A reductase activity in human fibroblasts by lipoproteins. Proc. Natl. Acad. U.S.A. *70*, 2162–2166.

17. Brown, M.S., and Goldstein, J.L. (1974). Familial hypercholesterolemia: Defective binding of lipoproteins to cultured fibroblasts associated with impaired regulation of 3-hydroxy-3-methylglutaryl coenzyme a reductase activity. PNAS *71*, 788–792.

18. Goldstein, J.L., and Brown, M.S. (1973). Familial hypercholesterolemia: identification of a defect in the regulation of 3-hydroxy-3-methylglutaryl coenzyme A reductase activity associated with overproduction of cholesterol. PNAS *70*, 2804–2808.

19. Hershko, A., Ciechanover, A., and Rose, I.A. (1979). Resolution of the ATP-dependent proteolytic system from reticulocytes: a component that interacts with ATP. Proc Natl Acad Sci U S A *76*, 3107–3110.

20. Ciechanover, A., Heller, H., Elias, S., Haas, A.L., and Hershko, A. (1980). ATP-dependent conjugation of reticulocyte proteins with the polypeptide required for protein degradation. Proc Natl Acad Sci U S A *77*, 1365–1368.

21. Hershko, A., Ciechanover, A., Heller, H., Haas, A.L., and Rose, I.A. (1980). Proposed role of ATP in protein breakdown: conjugation of protein with multiple chains of the polypeptide of ATP-dependent proteolysis. Proc Natl Acad Sci U S A *77*, 1783–1786.

22. Vale, R.D., Schnapp, B.J., Reese, T.S., and Sheetz, M.P. (1985). Movement of organelles along filaments dissociated from the axoplasm of the squid giant axon. Cell *40*, 449–454.

23. Schnapp, B.J., Vale, R.D., Sheetz, M.P., and Reese, T.S. (1985). Single microtubules from squid axoplasm support bidirectional movement of organelles. Cell *40*, 455–462.

24. Vale, R.D., Schnapp, B.J., Reese, T.S., and Sheetz, M.P. (1985). Movement of organelles along filaments dissociated from the axoplasm of the squid giant axon. Cell *40*, 449–454.

25. Vale, R.D., Reese, T.S., and Sheetz, M.P. (1985). Identification of a novel force-generating protein, kinesin, involved in microtubule-based motility. Cell *42*, 39–50.

26. Vale, R.D., Schnapp, B.J., Reese, T.S., and Sheetz, M.P. (1985). Organelle, bead, and microtubule translocations promoted by soluble factors from the squid giant axon. Cell *40*, 559–569.

27. Daniel, R.J. (2015). A generation at risk: young investigators and the future of the biomedical workforce. Proc. Natl. Acad. Sci. U S A 112: 313-318.

28. Polka, J.K., and Krukenberg, K.A. (2014). Making science a desirable career. Science *346*, 1422–1422.
21

**Figure 1**

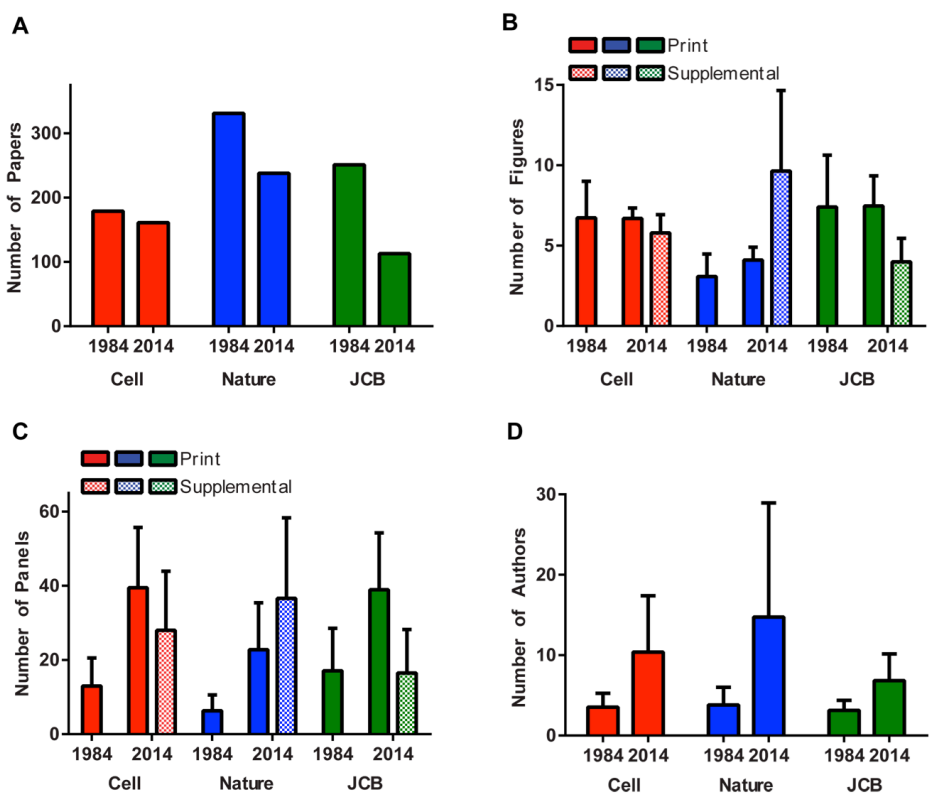

Fig. 1: Statistics for papers published in Cell, Nature (biology papers only) and the Journal of Cell Biology (JCB) for the months of January-June in 1984 and 2014. Long and short format papers (Articles and Letters for Nature, and Articles and Reports/Rapid Communications for JCB) are grouped together in this figure, but analysis of each category can be found in Fig. S1. A) The total number of papers published during these two six month time periods. B) The average number of figures in the print and online supplement of each paper. For Nature, most of the data in this figure is derived from the "Extended Data" section, although the "Supplemental Information" section also contributes some data in this analysis. An online supplement did not exist for journals in 1984. C) The number of panels per paper (assigned as a letter in the figure; tables were also scored in this category). D) The average number of authors per paper. The means and standard deviations are shown in panels B-D. See SI Methods for details on analysis. See Fig. S2 for an analysis of the pieces of distinct experimental data contained within the panels of the print versions of Cell and Nature.

**Figure 2**

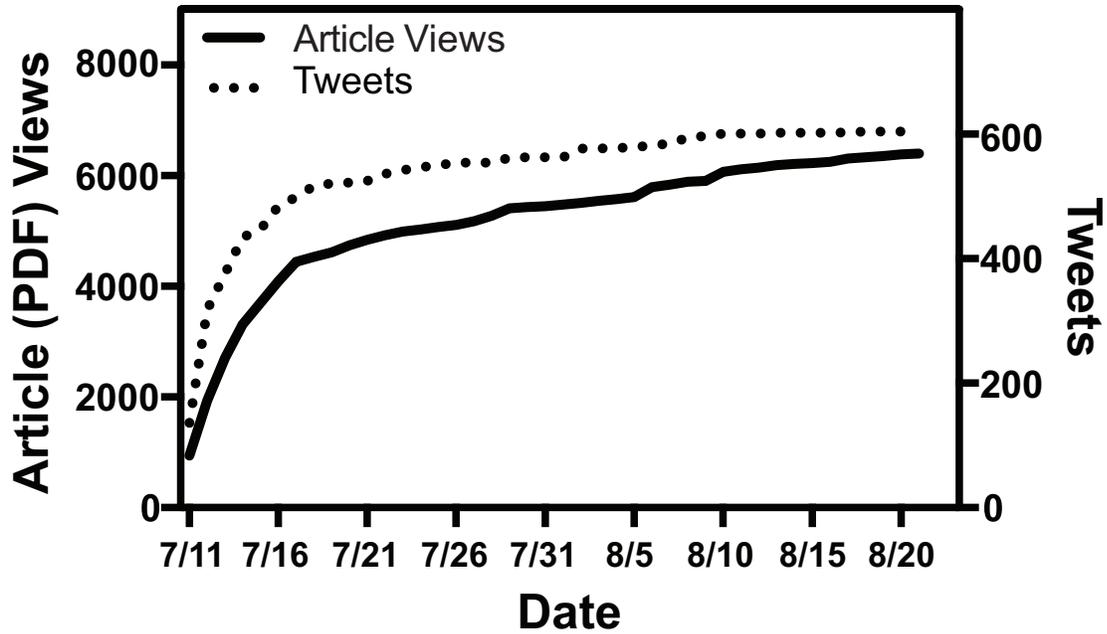

Fig. 2: Cumulative article (PDF) views and Tweets for the original version of this Perspective after its posting on bioRxiv (33). The data show the viewership and social media exchanges from the time of its posting (July 11, 2015) until the time when two peers reviews and a favorable editorial decision was transmitted to the author by PNAS (August, 21, 2015). Abstract views were more than twice the number of the PDF views. Information on daily views was provided by bioRxiv.

**Table 1: Scientific journal publications from UCSF graduate students.**

|  | No. Students | Grad Time (yrs) | Time to 1st author paper (yrs) | Number of 1st author publications | 1st+2nd author publications | 1st author C/N/S | 1st+2nd author C/N/S |
|---|---|---|---|---|---|---|---|
| 1979-89 | 71 | 5.7±1.0 | 4.7±2.3 | 2.2±1.5 | 2.9±1.8 | 0.52 | 0.80 |
| Top '79-89 | 24 | 5.2±0.9 | 3.4±1.1 | 3.1±1.2 | 4.5±1.7 | 1.25 | 1.63 |
| 2012-14 | 104 | 6.3±0.9 | 6.0±1.9 | 1.4±0.9 | 2.1±1.3 | 0.17 | 0.31 |
| Top '12-14 | 34 | 5.9±0.7 | 4.7±1.4 | 2.4±0.8 | 3.5±1.1 | 0.53 | 0.94 |

Table Footnote: The publications from Ph.D. students who performed experimental work and graduated in the indicated years of the Biochemistry and Molecular Biology, Biophysics, Genetics, and Neuroscience programs were analyzed. The time periods indicated refer to the year of graduation. A larger time span (1979-1980) was scored compared to the recent time period (2012-2014) since past graduate programs were smaller than they are now. "Top" refers to the top $1/3^{rd}$ of the students in each group with the best publication records, as assigned qualitatively based upon the combination of criteria described in this table. "C/N/S" refers to papers in Cell, Nature and Science and represents the average number of publications in these journals per student. Values represent means and standard deviations. Since co-authorship did not exist in the 1980s, we only scored the order of authorship; thus a shared first author in the second position was counted as a second authorship in our analysis; an exception to this rule was made if a second position, co-first author work was the sole paper from the student's graduate work. For more details of the analysis, see the SI Methods section. Scatter plots for all of the data are shown in Figs. S3 and S4.

# Supporting Information:

**SI Methods**

*Scoring of Panels and Data*

Panels were scored by simply counting the lettering (*a, b*, etc) designations in figures. Data-containing tables and figure schematics were counted as panels. Videos in the supplemental material were not counted. Panels are an imprecise proxy for the experimental data contained within a paper, and we therefore we attempted to estimate the amount of distinct pieces of data in Fig. S2. For example, a single experiment may be displayed in multiple panels with separate letters, such as different views of a fluorescence micrograph. Conversely, a single labeled panel may contact multiple different types of experiments. Therefore panels were scored as to whether they contained distinct pieces of data. To provide examples, if a representative image in one panel and quantification of the same experiment was provided in another panel, then both panels would be counted as a single piece of data. Also, if the same experiment was quantified in multiple ways (e.g. analysis of different organelle sizes or multiple kinetic parameters from the same experiment) and presented in multiple panels, then it would still be counted as a single piece of data. Different views or slices of the same sample, views of the same crystal structure, and multiple probes (for DNA or protein) used for the same sample also were considered as a single piece of information. Identical experiments applied to two different cell lines were also considered as one piece of data. Sequence alignments were counted as a one piece of data as were tables. Differentiation of separate pieces of data only were evaluated and scored between panels in a single figure and not between figures. Schematics and model figures were also not counted as "data" in this analysis. Two graduate students independently quantified the data presented in January and February 1984 articles in Cell to determine whether these criteria led to consistent scoring. The average pieces of distinct data per article were 7.33 and 7.16, indicating good overall agreement between two independent scorers. The other months of Jan-June from 1984 and 2014 for Cell and Nature were scored by a single person.

*Analysis of UCSF Graduate Student Publications*

Several basic science graduate programs in the 1980s have disappeared or merged with other programs and new graduate programs have formed more recently. To make a fair comparison of graduate student work between the 1980s and current times, we analyzed student data from four basic science PhD degree granting programs that have spanned both time periods: Biochemistry and Molecular Biology, Biophysics, Genetics, and Neuroscience. Since this study was focused on experimental science, students conducting exclusively theory or modeling studies were not counted in this analysis (5 students in 2012-4 in this category). Information on the time of entering graduate school and the time at which the degree was granted was obtained from the UCSF student registrar's office. Publication references and dates for the students were obtained by searching PubMed. Reviews or methods papers that were largely more detailed descriptions of previously published methods were not counted. "Shared authorship" represents a difficult issue, since this designation did not exist in the 1980s. While acknowledging the drawbacks of doing so, we only scored the order of authorship; thus a shared first author in the second position was counted as a second authorship in our analysis. The reason for doing so is to allow a more direct comparison with data from the 1980s, which did not employ co-first or co-second authorship as a credit sharing strategy. However, an exception was made for students that only published a single co-first author in their graduate work; in this case, this second-position work was counted as a first-author paper (6 student in this category). A second complication was scoring papers that were published a year or more after a degree was awarded. We directly emailed faculty or students from the 1980s to inquire whether such late publications were a product of their thesis work or primarily from a subsequent postdoctoral period (which were not scored). With only a couple of exceptions, these late publications were from thesis work; in many cases, difficulties in communication after leaving the laboratory between student and PI in the "pre-internet" era was cited as reasons for the delay in publication. However, papers published ~2 years beyond their graduation date were not scored in our analysis, unless it was their sole paper (1 student). For the recent UCSF graduate students, we contacted the PIs of students who graduated between June 2013-December 2014 to inquire whether the student was working on additional first or second author publications and whether the paper was in preparation, submission, revision, or in press. We added all anticipated publications to the student's data profile (17 students), estimating an approximate, best circumstance time of publication based upon the status described by the PI (~9 months for in preparation, 6 months for submitted, and 3 months for revision). It is possible that some of these anticipated papers may not be published or published with a longer time frame. If a student did not produce a first or a first/second author publication, then a "0" was entered for that category of publications. In the 1979-1989 group, there were 8 students without a first author publication and 4 students for whom we could not find a record of any publication in PubMed, although supporting evidence on the internet confirmed that they graduated. In the 2013-14 group, there were 9 students without an anticipated first author publication and 4 students without an anticipated first/second author publication. Students who did not publish a first-author paper were not included in the analysis of time to first author publication.

## Supporting Information Figures

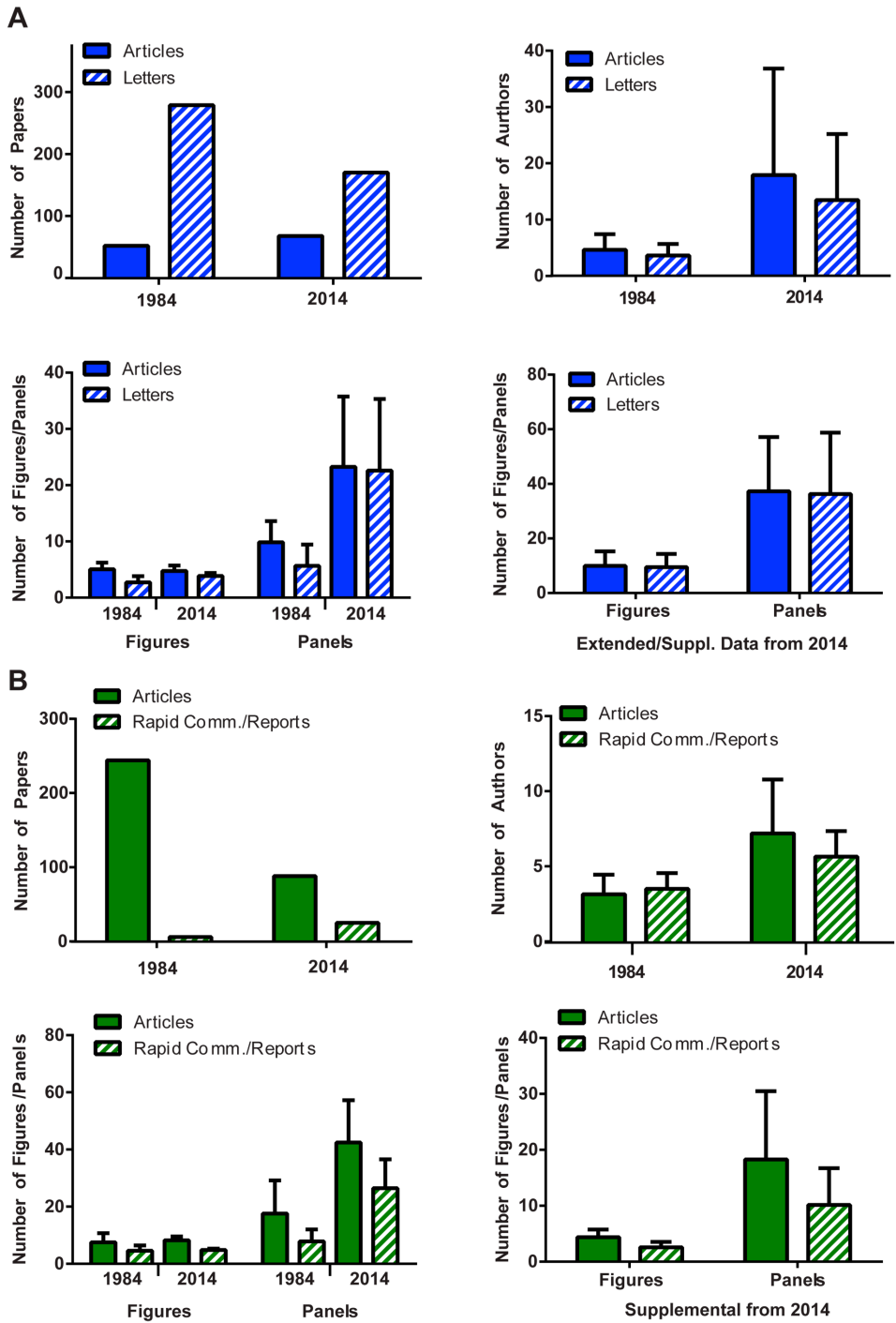

**Fig. S1:** Breakdown of information for long and short format papers. A) Data for Nature: long format (Articles) and short format (Letters). B) Data for Journal of Cell Biology (JCB): long format (Articles) and short format (Rapid Communications (1984 name) or Reports (2014 name). These data from long and short format papers were combined together in the analysis in Fig. 1.

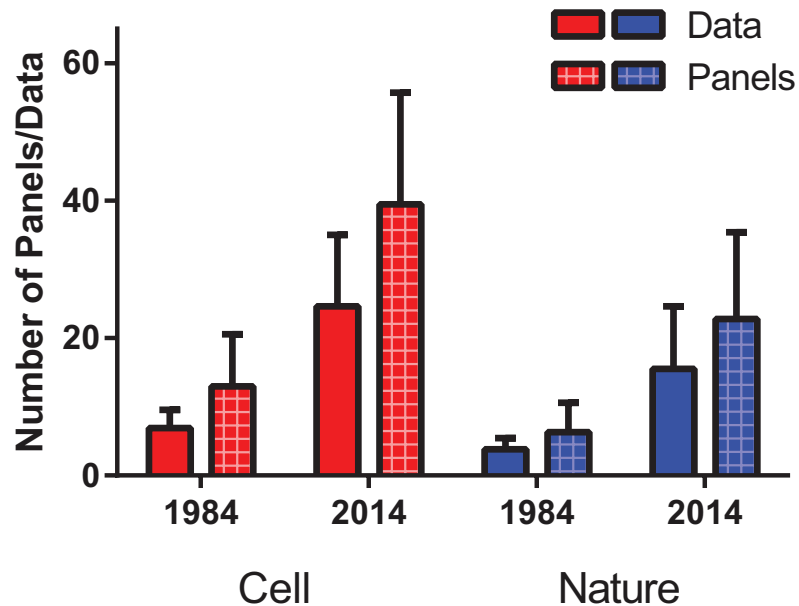

**Fig. S2:** Analysis of the number of panels (assigned as a letter in the figure) and distinct pieces of experimental data in the print versions of Cell and Nature. "Data" is defined as derived from a distinct experiment or a significant type of new analysis (see SI Methods section); as an example, two panels that show two views of a micrograph would be considered as a single datum in this analysis. While the scoring of "distinct data" is admittedly subjective, the analysis shows an approximately similar ratio of data versus panels in the two journals and between the two different time periods.

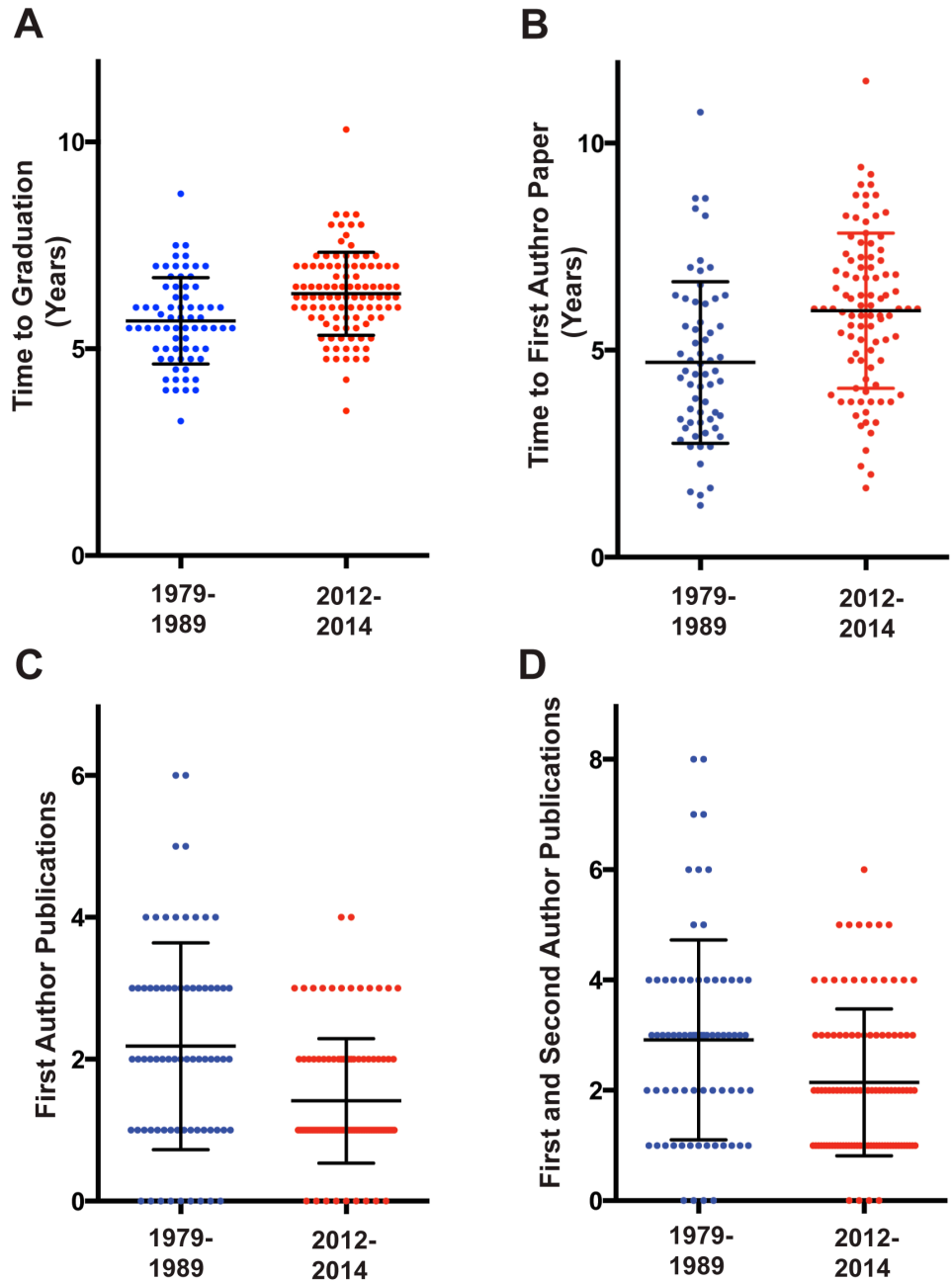

**Fig. S3:** Scatter plot of data on UCSF graduate students corresponding to Table 1. The time periods of graduation are indicated on the X axis (n =71 for 1979-1989 graduates; n = 104 for 2012-2014 graduates). The middle black lines indicate the mean and the error bars show standard deviations. Data for graduation and publication times were rounded to the nearest quarter of a year in this graph. The p-value differences (Kolmogorv-Smirnov test) for time to graduation, time to the first first-author publication, number of first-author publications, and number of first- and second-author publications are 0.0007, 0.0002, 0.0009, and 0.0083.

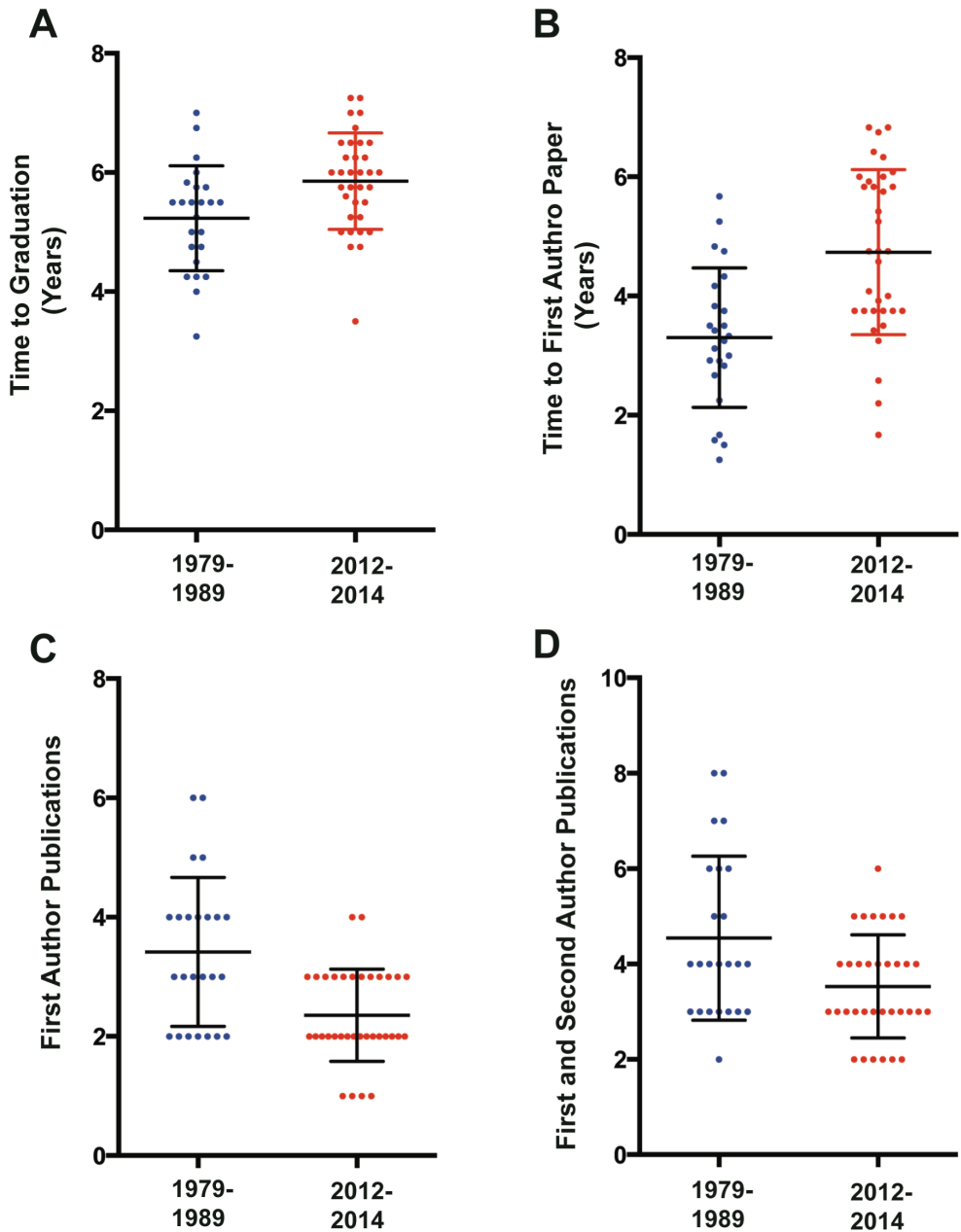

**Fig. S4:** Scatter plot of data of the top one-third UCSF graduate student group with the best publication record corresponding to Table 1. The time periods of graduation are indicated on the X axis (n =24 for 1979-1989 graduates; n = 34 for 2012-2014 graduates). The middle black lines indicate the mean and the error bars show standard deviations. Data for graduation and publication times were rounded to the nearest quarter of a year in this graph. The p-value differences (Kolmogorv-Smirnov test) for time to graduation, time to first first-author publication, number of first-author publications, and number of first- and second-author publications are 0.03, 0.002, 0.022, and 0.289.

**Q&A Regarding Pre-prints**

The following questions or concerns (paraphrased here in italics) were raised by others in response to the initial posting of this article on bioRxiv. My responses are presented below each question.

**<u>Reproducibility and Quality</u>**

*We already have a looming problem of irreproducibility. Pre-prints will just encourage more irreproducible results to be spread throughout the community.*

This issue is indeed important, since pre-prints open up the possibility of wide-spread science communication prior to peer review. Pre-prints might allow work to be disseminated, before mistakes are caught by peer review and thus lead researchers down wrong tracks. On the other hand, many peer-reviewed articles have proven to be inaccurate, and there is no clear data indicating how successful peer review is in filtering out irreproducible, inaccurate or fraudulent data. It might be better to have many people see the work right away, allowing the possibility of inadvertent mistakes to be caught and helping peer reviewers and the authors themselves to produce an accurate final product. Furthermore, a high profile result will likely be replicated right away and thus validated before it is published in a high profile journal. A good commenting system on pre-prints might help this process.

The immediate exposure of preprints also will likely be a motivating factor for accuracy. Many researchers intentionally do not complete all of their experiments in their first journal submission, since the journals emphasize "impact" in their first round of screening. Thus, mistakes in an initial journal submission and peer review are "invisible" and have no or minimal negative consequences for the author if the paper is rejected. This contrasts with a pre-print submission, in which all of the data is immediately transparent to the science community. This transparency will cause good scientists to be very cautious about their submission to a pre-print server, since that work will be seen and judged by their peers immediately. <u>Having the scientist decide when his/her work is ready for dissemination will be an empowering action and also one filled with a sense of responsibility.</u>

The subject of reproducibility, however, is a very complex one and should be taken into careful consideration. I would recommend collecting data on how pre-prints impact scientific reproducibility, but would argue that the disincentives for pre-prints (described in this Perpectives) should be removed first to allow increased use.

*Journal filters are good. I don't have time to sort through work in a massive pre-print server. I also am more assured of quality if I read work in top journals.*

Pre-prints will not replace the journals and instead will exist alongside them. You might prefer reading journals in order to learn about a new field, where the speed of access to new information might be less important. However, pre-prints would allow you to access to information faster in your own field, which might help to advance your research program. Thus, pre-prints and journal articles together can serve different needs in the scientific community. As discussed in the Perspective, it is also possible to experiment with filters that will allow users to sort through the content of pre-print servers for benchmarks of quality (e.g. specific scientists, the funding source that supported the work, recommendations from user groups, etc).

*There are already many low interest, low quality papers being published in journals. Won't a pre-print server just accentuate this problem and further plague our scientific community?*
Most scientists seek to establish a good reputation and thus will want to showcase high quality work to their colleagues, regardless of whether it is through a pre-print or journal. Some scientific material that is currently hard to publish, such as confirming a finding or reporting a negative result, might be posted on pre-print servers, thus adding more scientific material than is currently being accepted at journals. However, as discussed above, the best solution will be to create better mechanisms of searching for relevant information that appears both in pre-prints and in journals.

## **Journals and Pre-Prints**

*With potential comments being posted on pre-prints, won't this endanger the subsequent journal-based peer review process?*
This will have to be tested in practice. arXiv does not have a comment system but bioRxiv does. One might argue that commenting could improve subsequent peer review, if thoughtful people use the commenting system. For example, a particularly good comment on a pre-print could help a journal referee in their review. Importantly, because the identity of the pre-print commenter is known, the system will prevent competitors from making negative remarks behind a cloak of anonymity. Furthermore, through a pre-print, authors can receive direct feedback on their work from the community. Such comments, some of which might not have arisen through journal peer review, can help the author to revise their work and publish a better paper in the end. Thus, pre-prints could facilitate direct feedback to authors and information for referees, both of which could lead to improved revisions of the work.

*Someone posts a pre-print with a quick and dirty experiment to make a claim. I worked much harder to establish proof with a more complete and convincing set of evidence. I am now forced to post my pre-print a month later. Won't journals be reluctant to publish my paper since they will have seen the earlier posted work?*
Quite the opposite may occur. Currently journals want to publish stories first, but some of this drive may diminish if work routinely appears first as pre-prints. Journals then may be incentivized to look more towards quality than speed and seek to <u>publish the definitive work</u> that will stand the test of time and become the publication that is most cited. Furthermore the issue of speed versus quality of research already exists in the present journal system. For example, a researcher can quickly publish a study with minimal data in a lower journal; this publication can potentially color another journal's view of a more extensive manuscript being submitted later. Furthermore, if a scientist repeatedly has a pattern of reporting quick and dirty experiments to beat competitors rather than doing complete and thoughtful work, then this will tarnish his/her reputation and will not be a path to long term success. In addition, there is "version control" with pre-prints; if someone rushes out an incomplete paper and then subsequently wants to correct mistakes, they can upload a new version, but the original version remains on the site for all to see.

*How are news and publicity handled if there is a preprint submission as well as a subsequent journal publication*

Historical examples from arXiv reveal various ways in which this has been handled. In some cases, a journal or press will "find" a pre-print on arXiv and run a story on the work prior to journal publication. In some cases, the pre-print will be posted on arXiv at the time of acceptance to a journal (but prior to publication), and the press will cite the arXiv pre-print and name the journal in which it will be ultimately published. Even government agencies such as NSF have issued publicity surrounding a pre-print. In other cases, publicity only arises with the greater attention associated with the journal publication. A critical issue is that the authors need to follow the embargo policy of the journal to which they intend to submit, which usually prohibits the authors from speaking about their work directly with the press themselves prior to publication. See Nature's guidelines on publicity": http://www.nature.com/authors/policies/confidentiality.html. In general, news and publicity has been managed successfully with arXiv and the journal system.

*A bigger issue to me is open access.*

Pre-prints are free for anyone in the world. Use of this system will therefore ensure that there is always a version of manuscript that is freely available, regardless of what journal it is eventually published in. However, for certain journals, the accepted version of an article cannot be posted as pre-print for up to six months from the time of publication (e.g. see Nature guidelines cited above).

*Having pre-prints listed on PubMed would be helpful as one-stop shopping to find science content*

Currently PubMed is only for peer-reviewed articles. To facilitate content discovery, one could imagine developing a new biologist-friendly search engine that will search for content on PubMed, bioRxiv and arXiv. On the other hand, such functions could be integrated into PubMed. Both solutions are workable, and the community and NIH can decide on the best course of action.

**Ethical and Practical Issues for Biology**

*Experimental biology is moving so fast. I am worried that if I post on bioRxiv or arXiv then someone will scoop me by rushing a paper to a journal and perhaps be luckier in the publication process.*

The possibility that results/ideas might be "stolen" from a pre-print, resulting in the loss of credit for researcher, seems to be a prevalent concern in the biology community. This is why some argue that pre-prints simply will not work in biology as they have in physics. Here is an excerpt from a reviewer's comment on this Perspective from PNAS:

"Should the author choose to continue to push the prepublication format, he might anticipate the following criticism of his logic. He poses that prepublication works for experimental physics so it can work for experimental biology. This analogy appears flawed. Physics today is like biology 40 years ago. The experimental systems needed to address a problem are unique, for example a synchrotron to address a problem in subatomic physics (like a bicoid mutant that nobody but Ed Lewis had). Hence, a prepublication is safe. Nobody can quickly generate the data of the prepublication or has preliminary data similar to the prepublication. What makes current biology so exciting is the lightening fast connections that are made between very rapidly moving systems. These same connections generate problems for the prepublication concept. Here is the scenario that critics will bring forward. One has a very nice unpublished discovery and talks about it at a meeting or University. A member of the audience has some preliminary results in another system that in the context of the talk all of sudden make sense. With much greater confidence the member of the audience adds a few experiments, publishes these results and common conclusion in a prepublication. This minimal publication is much weaker than the lecture but nonetheless gets priority. This scenario can't or would rarely happen in physics but would be the fear of every biologist talking about unpublished results. Nobody would share unpublished results because the speed at which unrefereed results could be published. In this light the author might use the text to probe a little deeper why biology did not move to prepublication format if in fact biology and physics are interchangeable. As it is, he will get much criticism for comparing apples to oranges."

These remarks are thoughtful and reflect many people's concerns. My response is that not all biology experiments are so lightening fast to repeat. Some are, but most papers are fairly complex and not trivial to repeat in a few weeks even by a well-established competing lab. However, talking about work in a lecture constitutes a problem for establishing priority, as the referee indicates. Physicists tend to acknowledge information transmitted in public talks. But part of the motivation for establishing arXiv was to create a common access point where a discovery could be announced to the community, since not everyone can attend a lecture. Because arXiv is so widely viewed by the community, it is very difficult for an individual to "steal and run" with an idea/experiment with the excuse that they never saw it on arXiv. If pre-prints are going to be successful, they must carry with them the gravitas of priority. Finally, I asked Paul Ginsparg, founder of arXiv, if there were examples of transgressions where a result was posted on arXiv and then someone copied it and published it faster in a journal to claim priority. He could not think of a single example where this happened and also thought that the physics community would not tolerate such behavior. They also would not tolerate someone publishing a cheap paper on arXiv in response to hearing an outstanding work or idea in a public lecture. Furthermore, work appearing close in time as pre-prints (e.g. within a couple of months) will be compared based upon quality and acknowledged as co-discoveries if they deserve to be, just as is the case with journal publications. Perhaps physicists are not behind biologists (see the referee's comment), but rather are 40 years ahead of us in science communication and ethics. The interesting question is how does one define priority and associated ethical practices in biology? Perhaps the signing of a declaration by leaders in the biology might be helpful in initiating the process and setting a new tone. Ultimately, however, it will have to be further propagated by investigators themselves and how they teach their trainees.

*Biologists develop specialized reagents and strains for their work; there is an obligation to release these reagents immediately to the community upon publication. Will this obligation apply to preprint postings?*
       Some investigators may be happy to release their reagents or share software at the pre-print stage. Others may be reluctant to do so until after journal publication, especially given current concerns described above. Thus, the community may wish to develop a

policy once pre-prints become more widely used. One could imagine a grass-roots agreement providing a grace period (for example, 1 year) by the end of which all reagents, strains, and all source data must be made publicly available after a preprint is posted. Any such recommendation policy could be reevaluated as the system matured.

*The main problem in the life sciences is the lack of academic and industry jobs and excessive competition for those jobs. Getting my work out earlier with a pre-print is not going to help me get a job, especially if everyone is posting pre-prints.*
   Agreed. Posting pre-prints will not help you get a job *per se*, since that is determined by competition with other applicants. However, a pre-print might be useful in some circumstances, since it will allow a potential employer to access your work if it has not yet been published or held up in a prolonged review. Currently, a manuscript listed as "submitted" on a CV counts for very little. Furthermore, in physics, recent pre-prints on arXiv play a crucial role in evaluating candidates for jobs.

*The love of just a few elite journals is the biggest problem in life sciences these days. I don't see how the pre-prints are going to solve this issue.*
   Pre-prints will not truly solve this issue. However, they might represent the start of a longer-term change in how scientific work is evaluated. Because of the necessity to stay informed, scientists will read pre-prints in their own field and make judgments of the quality before it has a journal name attached to it. Grant reviewers might also start to comment on the quality of work posted as a pre-print if it is presented as key evidence for a new research program. For such a vision to succeed, the best work in biology needs to be posted on a pre-print server and not solely routed to the elite journals. Leaders in the biomedical community will have to post their best work as pre-prints to set an example.

*What about medical sciences? If a pre-print on a medical procedure or a drug is posted but is wrong, then it might have disastrous consequences on patient care.*
   This is a reasonable question, especially given existing concerns on irreproducible work in medically-related areas being published in peer review journals. The medical sciences community will have to confront this issue themselves and decide on the best path. Biology is not a single monolithic enterprise but is composed of many different disciplines and communities. These different communities can decide when or if pre-prints represent a good mechanism of communicating their results. Following the history of arXiv, different communities (e.g. different branches of physics, mathematics and computational sciences) embraced pre-prints at different times.

**Feasibility of pre-prints and the potential of other mechanisms**

*Scientists are set in their ways. No one is going to use pre-prints.*
   Scientists are indeed conservative with regard to their habits. They are also unlikely to change their habits simply based on altruism. However, they will use pre-prints if they provide practical benefits for their research and careers. Pre-prints could benefit scientists if they: 1) allow them to establish priority for a discovery in a more predictable way than navigating an unpredictable journal review process, 2) allow them to use pre-prints as evidence of productivity in grant applications, particularly in cases where a new research direction is being pursued, 3) enable grad students to provide evidence of scholarly work for graduation or post-doc applications, thereby potentially decreasing their training time by many months, and 4) allow scientists to obtain feedback on their work earlier than is currently possible through the journal system.  While it is unlikely that everyone will switch to pre-prints, its use might increase significantly if people try it and have good experiences (as has occurred with arXiv).  Also, younger scientists may be "less set in their ways" than more senior scientists.  They have grown up socializing in an internet world, so the notion of sharing information through a pre-print server will not seem so foreign to them.

*I like the idea of pre-prints, but I hesitate to advise junior faculty in my department to submit pre-prints as it might not be good for their career.*

Here is a recent experience from James Fraser, a junior faculty member at UCSF:
"We submitted the paper (http://www.ncbi.nlm.nih.gov/pubmed/26280328) and the preprint (http://biorxiv.org/content/early/2015/02/03/014738) in February. In the intervening months before the paper was published online in August (publication went smoothly, with a supportive editor and constructive reviews), the following events happened based upon the information made available through the pre-print:
* Our software was downloaded by multiple groups around the world and used locally at UCSF to improve other EM structures.
* I was invited to an EM validation meeting to discuss the work (even though we hadn't published in that area before).
* My student was invited to speak at the local bay area EM meeting.
* My student got a fellowship (ARCS). He probably would have gotten it anyway - but having a biorxiv doi to point to for his "in review" paper may have helped.
* I talked to people about the method at multiple meetings and was able to point them to the preprint to judge for themselves."

*bioRxiv has been around since 2013 and it has a small following.  Hasn't the experiment been done already and the answer is in hand?*

I would argue that the experiment has not been done properly. Currently there are several major disincentives for preprints, which include: 1) an inability to cite a bioRxiv pre-print on NIH grants, 2) possible restrictions in subsequently publishing the work in certain journals, and 3) the potential of being "scooped" since it is unclear as to whether a pre-print constitutes "priority" amongst scientific peers for a discovery.  Note- "priority" among peers is a culture issue of assigning credit within the profession and differs from the legal term of "disclosure" (e.g. for a patent), which involves any public presentation. Given these restrictions, it is difficult to strongly recommend pre-prints in their current state. These deterrents need to be removed in order to give pre-prints a fair chance.

*There are better ways of transmitting scientific information than a pre-print plus journal system.  Why stall the inevitable by encouraging pre-prints?  Shouldn't we build a completely new system that will replace both journals and pre-prints?*

Scientific communication will likely evolve in new ways over the coming years and decades.  The question is how to get from where we are now to where we want to be. Replacing the current journal system now with something new is likely to meet considerable resistance and thus likely fail. Pre-prints, on the other hand, represent a viable evolutionary intermediate. Pre-prints can co-exist with the journal system, and thus do not represent an either-or choice for scientists. Also, supporting pre-prints should not prevent other desirable changes in the science communication system that our community would like to establish later on (e.g. changes in pre- or post-publication review and evaluation). Indeed, a short-term success with pre-prints would convey a message to our community that we are not locked into the status quo and that other changes are possible over time.

*F1000 R has a complete publishing platform that communicates the "pre-print" but also initiates transparent peer review and then indexes successfully peer reviewed papers on PubMed. What about systems like this?*

F1000 R has an interesting and new publishing process. However, submitting a work to F1000 R precludes submitting the same work in another journal (unlike bioRxiv or arXiv). Individual scientists will have to decide on a publishing mechanism that makes sense for them- submission to F1000 R, or through a pre-print server (bioRxiv/arXiv) plus a subsequent journal of their choice, or through other mechanisms.

*Are you sure that pre-prints will work in biology?*

No. I am only sure of death and taxes. However, we have to try experiments in scientific communication. This seems to be a relatively easy one to try, since the potential harm, cost, and infrastructure are minimal and the current barriers are not so difficult to overcome. If pre-prints are tried but do not succeed, then the answer will be clear and new ideas can be investigated.